\documentclass[reqno]{amsart}

\newcommand{\Rm}{\mathbb{R}}
\newcommand{\Cm}{\mathbb{C}}

\newcommand{\Sm}{\mathbb{S}}
\newcommand{\ba}{\begin{eqnarray*}}
\newcommand{\ea}{\end{eqnarray*}}
\newcommand{\be}{\begin{equation}}
\newcommand{\ee}{\end{equation}}
\newcommand{\bea}{\begin{eqnarray}}
\newcommand{\eea}{\end{eqnarray}}
\newcommand{\va}{\varphi}

\newcommand{\vv}[1]{\boldsymbol{\mathrm{#1}}}
\newcommand{\hvv}[1]{\boldsymbol{\hat{\mathrm{#1}}}}
\newcommand{\uv}{\boldsymbol{{\hat{\Omega}}}}
\newcommand{\uvk}{\boldsymbol{\hat{\mathrm{k}}}}

\newcommand{\rrf}[1]{\mathop{\mathcal{R}_{{#1}}}}

\usepackage{amsmath}
\usepackage{amsthm,amssymb,mathrsfs}

\theoremstyle{remark}

\title[]{The Green's function for the three-dimensional linear Boltzmann 
equation via Fourier transform}

\author[]{Manabu Machida}

\begin{document}

\begin{abstract}
The linear Boltzmann equation with constant coefficients in the 
three-dimensional infinite space is revisited. It is known that the Green's 
function can be calculated via the Fourier transform in the case of isotropic 
scattering. In this paper, we show that the three-dimensional Green's 
function can be computed with the Fourier transform even in the case of 
arbitrary anisotropic scattering.
\end{abstract}

\maketitle

\section{Introduction}

We consider the linear Boltzmann equation in three dimensions, which governs 
neutron transport and radiative transfer. If scattering is isotropic, 
it is well known that the Green's function of the monoenergetic neutron 
transport in a three-dimensional infinite medium can be obtained using 
the Fourier transform \cite{Case-Zweifel,Ishimaru78}. In one dimension, 
Ganapol developed Fourier transform techniques and showed that the Green's 
function can be found even for arbitrary anisotropic scattering 
\cite{Ganapol00,Ganapol15}. In this paper, we will extend Ganapol's 
calculation in one-dimensional transport theory to three dimensions 
making use of rotated reference frames and present the three-dimensional 
Green's function for arbitrary anisotropic scattering.

The introduction of rotated reference frames in neutron transport theory 
goes back to Dede \cite{Dede64} and Kobayashi \cite{Kobayashi77}. Dede 
discussed that three-dimensional equations in the $P_N$ method reduce to 
one-dimensional equations by measuring angles in the reference frame rotated 
in the direction of the Fourier vector. Kobayashi's work is similar to 
the calculation in the present paper in the sense that the recurrence relation 
(\ref{Eq10a}) was derived, however, $\bar{\psi}_l^m$ was not explicitly 
obtained as we will do in (\ref{Eq22a}). The first practical way of using 
rotated reference frames, which made numerical calculation possible, was 
found by Markel \cite{Markel04}. Markel established an efficient method of 
computing the specific intensity of light in three dimensions by expressing 
the specific intensity in terms of eigenmodes and rotating the reference 
frame for each eigenmode \cite{Markel04,Panasyuk06}. The technique 
was also applied to inverse transport problems \cite{Schotland-Markel07}. 
Recently, it was found that the use of such rotated reference frames is not 
restricted to Legendre polynomials and spherical harmonics. Case's 
singular eigenfunctions were extended to three dimensions \cite{Machida14}. 
With this result, the $F_N$ method \cite{Siewert78} was extended to 
three dimensions \cite{Machida15}.

Let us write the Green's function of monoenergetic neutron transport in 
three dimensions. The angular flux $G(\vv{r},\uv;\uv_0)\in\Rm$ 
($\vv{r}\in\Rm^3$, $\uv\in\Sm^2$) obeys
\[
\left(\uv\cdot\nabla+1\right)G(\vv{r},\uv;\uv_0)
=c\int_{\Sm^2}p(\uv,\uv')G(\vv{r},\uv';\uv_0)\,d\uv'
+\delta(\vv{r})\delta(\uv-\uv_0),
\]
where $c$ ($0<c<1$) is the albedo for single scattering and 
$p(\uv,\uv')\in\Rm$ is the phase function. The unit vector $\uv$ has 
the polar angle $\theta$ and azimuthal angle $\va$. The source placed at the 
origin $\vv{r}=\vv{0}$ emits neutrons in the direction $\uv_0$. We assume 
$p(\uv,\uv')$ depends only on $\uv\cdot\uv'$ and write
\[
p(\uv,\uv')
=\sum_{l=0}^L\sum_{m=-l}^l\frac{\beta_l}{2l+1}Y_{lm}(\uv)Y_{lm}^*(\uv')
=\frac{1}{4\pi}\sum_{l=0}^L\sum_{m=-l}^l\omega_l^mP_l^m(\mu)P_l^m(\mu')
e^{im(\va-\va')},
\]
where $\beta_0=1$, $|\beta_l|<2l+1$ ($l=1,2,\dots,L$), and we defined 
$\mu=\cos\theta$ and
\[
\omega_l^m=\beta_l\frac{(l-m)!}{(l+m)!}.
\]
Here $Y_{lm}(\uv)$ are spherical harmonics given by
\[
Y_{lm}(\uv)=\sqrt{\frac{2l+1}{4\pi}\frac{(l-m)!}{(l+m)!}}P_l^m(\mu)e^{im\va},
\]
and
\[
P_l^m(\mu)=(-1)^m(1-\mu^2)^{m/2}\frac{d^m}{d\mu^m}P_l(\mu),\quad
P_l^{-m}(\mu)=(-1)^m\frac{(l-m)!}{(l+m)!}P_l^m(\mu),\quad
0\le m\le l.
\]
Associated Legendre polynomials $P_l^m(\mu)$ satisfy the following 
recurrence relation.
\be
(2l+1)\mu P_l^m(\mu)=(l-m+1)P_{l+1}^m(\mu)+(l+m)P_{l-1}^m(\mu),
\label{recurrencePlm}
\ee
with initial terms
\[
P_m^m(\mu)=(-1)^m(2m-1)!!(1-\mu^2)^{m/2},\quad
P_{m+1}^m(\mu)=(2m+1)\mu P_m^m(\mu),\quad 0\le m\le l.
\]

The first two terms in the collision expansion of the Green's function contain 
the Dirac delta function. We subtract the uncollided part as follows. 
\[
G(\vv{r},\uv;\uv_0)=G_0(\vv{r},\uv;\uv_0)+\psi(\vv{r},\uv),
\]
where $G_0(\vv{r},\uv;\uv_0)$ satisfies
\[
\left(\uv\cdot\nabla+1\right)G_0(\vv{r},\uv;\uv_0)
=\delta(\uv-\uv_0)\delta(\vv{r}),
\]
and $\psi(\vv{r},\uv)$ satisfies
\[
\left(\uv\cdot\nabla+1\right)\psi(\vv{r},\uv)
=c\int_{\Sm^2}p(\uv,\uv')\psi(\vv{r},\uv')\,d\uv'
+\frac{c}{r^2}e^{-r}p(\uv,\uv_0)\delta(\hvv{r}-\uv_0),
\]
where
\[
r=|\vv{r}|,\qquad\hvv{r}=\frac{\vv{r}}{r}.
\]
The source term in the transport equation for $\psi(\vv{r},\uv)$ can be 
calculated by using
\[
G_0(\vv{r},\uv;\uv_0)=\frac{1}{r^2}e^{-r}\delta(\uv-\hvv{r})\delta(\uv-\uv_0).
\]
In this paper we will consider how $\psi(\vv{r},\uv)$ is obtained.

In the case of isotropic scattering ($L=0$), we can compute $\psi(\vv{r},\uv)$ 
with the textbook way (Appendix \ref{textbook}) as
\[
\psi(\vv{r},\uv)=\frac{c}{2(2\pi)^4}\int_{\Rm^3}e^{i\vv{k}\cdot\vv{r}}
\frac{\left[1-\frac{c}{k}\tan^{-1}(k)\right]^{-1}}
{(1+i\vv{k}\cdot\uv)(1+i\vv{k}\cdot\uv_0)}\,d\vv{k}.
\]
The aim of this paper is to extend this result to arbitrary anisotropic 
scattering.

As the first main result, we obtain $\psi(\vv{r},\uv)$ as
\be
\psi(\vv{r},\uv)
=\frac{c}{2(2\pi)^4}\int_{\Rm^3}e^{i\vv{k}\cdot\vv{r}}
\frac{M(\vv{k},\uv,\uv_0)}{(1+i\vv{k}\cdot\uv)(1+i\vv{k}\cdot\uv_0)}\,d\vv{k},
\label{mainmatrix}
\ee
where $M(\vv{k},\uv,\uv_0)$ is given in (\ref{defM}).

Since the calculation of $M(\vv{k},\uv,\uv_0)$ involves matrix inversion, 
we explore an alternative expression of $\psi(\vv{r},\uv)$. As the second main 
result, we will show that $\psi(\vv{r},\uv)$ is given by
\be
\psi(\vv{r},\uv)
=\frac{1}{(2\pi)^3}\sum_{l=0}^{\infty}\sum_{m=-l}^lY_{lm}(\uv)\int_{\Rm^3}
e^{i\vv{k}\cdot\vv{r}}e^{-im\va_{\uvk}}\kappa_{lm}(\vv{k})\,d\vv{k},
\label{mainresult}
\ee
where
\ba
\kappa_{lm}(\vv{k})
&=&
\sum_{m'=-l}^l\frac{1}{g_{|m'|}^{m'}(i/k)}
\sqrt{\frac{2l+1}{4\pi}\frac{(l-m')!}{(l+m')!}}e^{-im'\va_0}
\nonumber \\
&\times&
d_{mm'}^l(\theta_{\uvk})
\left[g_l^{m'}(\frac{i}{k})\bar{\psi}_{|m'|}^{m'}(\frac{i}{k},\hvv{k})
+\chi_l^{m'}(\frac{i}{k})\right].
\ea
Here, $d_{m'm}^l$ are Wigner's $d$-matrices \cite{Varshalovich} and $g_l^m$ 
are Chandrasekhar's polynomials of the first kind. Below, $\chi_l^m$ and 
$\bar{\psi}_{|m|}^m$ are given in (\ref{defchi}) and (\ref{Eq23a}), 
respectively.

Suppose that $\bar{\psi}_{|m|}^m$ is independent of $\va_{\uvk}$. We can 
write $\kappa_{lm}(\vv{k})=\kappa_{lm}(k,\mu_{\uvk})$. In this case we 
obtain
\bea
\psi(\vv{r},\uv)
&=&
\frac{1}{(2\pi)^2}\sum_{l=0}^{\infty}\sum_{m=-l}^lY_{lm}(\uv)i^m\int_0^{\infty}
k^2
\nonumber \\
&\times&
\int_{-1}^1J_m\left(kr\sqrt{1-\mu_{\uvk}^2}\sin\theta_{\hvv{r}}\right)
e^{ikr\mu_{\uvk}\cos\theta_{\hvv{r}}}e^{-im\va_{\hvv{r}}}
\kappa_{lm}(k,\mu_{\uvk})\,d\mu_{\uvk}dk,
\label{mainresult2}
\eea
where $J_m$ is the Bessel function of degree $m$. For example, 
$\kappa_{lm}(k,\uvk)$ is independent of $\va_{\uvk}$ if $\uv_0=\hvv{z}$.

In what follows, we derive (\ref{mainmatrix}) in \S\ref{fouriertransform} and 
(\ref{mainresult}) in \S\ref{nonstandard}. In \S\ref{numerics}, we compute 
the energy density by using (\ref{mainresult2}). 
The key idea of rotated reference frames is introduced in the next section.

\section{Rotated reference frames}
\label{rotatedreferenceframes}

We introduce the operator $\rrf{\uvk}:\Cm\mapsto\Cm$ for a unit vector 
$\uvk\in\Cm^3$ ($\uvk\cdot\uvk=1$). By operating $\rrf{\uvk}$ we measure 
$\uv$ in the reference frame whose $z$-axis lies in the direction of $\uvk$ 
\cite{Markel04}. For example, we have
\[
\uv\cdot\uvk=\rrf{\uvk}\mu.
\]
If a function $f(\uv)\in\Cm$ is given as
\[
f(\uv)=\sum_{l=0}^{\infty}\sum_{m=-l}^lf_{lm}Y_{lm}(\uv),
\]
we have
\[
\rrf{\uvk}f(\uv)=\sum_{l=0}^{\infty}\sum_{m=-l}^lf_{lm}\sum_{m'=-l}^l
e^{-im'\va_{\uvk}}d_{m'm}^l(\theta_{\uvk})Y_{lm'}(\uv),
\]
where $\theta_{\uvk}$ and $\va_{\uvk}$ are the polar and azimuthal angles of 
$\uvk$ in the laboratory frame. In particular, we have
\[
\rrf{\uvk}Y_{lm}(\uv)=\sum_{m'=-l}^l
e^{-im'\va_{\uvk}}d_{m'm}^l(\theta_{\uvk})Y_{lm'}(\uv).
\]

\section{Fourier transform}
\label{fouriertransform}

We begin by noting that 
$\uv\cdot\uv'=(\rrf{\uvk}\uv)\cdot(\rrf{\uvk}\uv')$ and 
\[
p(\uv,\uv')=\sum_{l=0}^L\sum_{m=-l}^l\frac{\beta_l}{2l+1}
\left(\rrf{\uvk}Y_{lm}(\uv)\right)\left(\rrf{\uvk}Y_{lm}^*(\uv')\right),
\]
for an arbitrary unit vector $\uvk$. The transport equation is written as
\ba
\left(\uv\cdot\nabla+1\right)G(\vv{r},\uv;\uv_0)
&=&
c\sum_{l=0}^L\sum_{m=-l}^l\frac{\beta_l}{2l+1}
\left[\rrf{\uvk}Y_{lm}(\uv)\right]
\int_{\Sm^2}\left[\rrf{\uvk}Y_{lm}^*(\uv')\right]G(\vv{r},\uv';\uv_0)\,d\uv'
\\
&+&
\delta(\vv{r})\delta(\uv-\uv_0),
\ea

We introduce
\[
G_l^m(\vv{r})=
\sqrt{\frac{4\pi}{2l+1}\frac{(l+m)!}{(l-m)!}}e^{im\va_0}
\int_{\Sm^2}\left[\rrf{\uvk}Y_{lm}^*(\uv)\right]G(\vv{r},\uv;\uv_0)\,d\uv.
\]
An arbitrary vector $\vv{k}\in\Rm^3$ is given by $k$ ($0\le k<\infty$) and 
$\uvk\in\Rm^3$ ($\uvk\cdot\uvk=1$) as
\[
\vv{k}=k\uvk.
\]
With this vector $\vv{k}$ we perform the Fourier transform as
\ba
\bar{G}(\vv{k},\uv;\uv_0)&=&
\int_{\Rm^3}e^{-i\vv{k}\cdot\vv{r}}G(\vv{r},\uv;\uv_0)\,d\vv{r},
\\
\bar{G}_l^m(\vv{k})&=&
\int_{\Rm^3}e^{-i\vv{k}\cdot\vv{r}}G_l^m(\vv{r})\,d\vv{r}.
\ea
In the Fourier space we obtain
\ba
\left(1+i\vv{k}\cdot\uv\right)\bar{G}(\vv{k},\uv)
&=&
c\sum_{l=0}^L\sum_{m=-l}^l\frac{\omega_l^m}
{\sqrt{4\pi(2l+1)}}\sqrt{\frac{(l+m)!}{(l-m)!}}e^{-im\va_0}
\left(\rrf{\uvk}Y_{lm}(\uv)\right)\bar{G}_l^m(\vv{k})
\\
&+&
\delta(\uv-\uv_0).
\ea
This is expressed as
\bea
\bar{G}(\vv{k},\uv)
&=&
\frac{c}{2}\frac{1}{1+i\vv{k}\cdot\uv}\sum_{l=0}^L\sum_{m=-l}^l
\frac{\omega_l^m}{\sqrt{\pi(2l+1)}}\sqrt{\frac{(l+m)!}{(l-m)!}}e^{-im\va_0}
\left(\rrf{\uvk}Y_{lm}(\uv)\right)\bar{G}_l^m(\vv{k})
\nonumber \\
&+&
\frac{1}{1+i\vv{k}\cdot\uv}\delta(\uv-\uv_0).
\label{Eq4a}
\eea
By multiplying $\rrf{\uvk}Y_{lm}^*(\uv)$ on both sides of (\ref{Eq4a}) and 
integrating over $\Sm^2$, we obtain
\ba
\int_{\Sm^2}\left[\rrf{\uvk}Y_{lm}^*(\uv)\right]\bar{G}(\vv{k},\uv)\,d\uv
&=&
\frac{c}{2}\sum_{l'=0}^L\sum_{m'=-l'}^{l'}\frac{\omega_{l'}^{m'}}
{\sqrt{\pi(2l'+1)}}\sqrt{\frac{(l'+m')!}{(l'-m')!}}e^{-im'\va_0}
\bar{G}_{l'}^{m'}(\vv{k})
\\
&\times&
\int_{\Sm^2}\frac{1}{1+i\vv{k}\cdot\uv}\left[\rrf{\uvk}Y_{lm}^*(\uv)\right]
\left[\rrf{\uvk}Y_{l'm'}(\uv)\right]\,d\uv
\\
&+&
\frac{\rrf{\uvk}Y_{lm}^*(\uv_0)}{1+i\vv{k}\cdot\uv_0}.
\ea
We note that
\[
\vv{k}\cdot\uv=k\rrf{\uvk}\mu.
\]
Let us define
\[
L_{jl}^m(z)=\frac{z}{2}\int_{-1}^1\frac{P_j^m(\mu)P_l^m(\mu)}{z-\mu}\,d\mu,
\]
where $z=-1/(ik)$. Let $\Theta(\cdot)$ be the step function such that 
$\Theta(x)=1$ for $x\ge0$ and $\Theta(x)=0$ otherwise. We have
\be
\bar{G}_j^m(\vv{k})
=\Theta(L-|m|)c\sum_{l=|m|}^L\omega_l^mL_{jl}^m(z)\bar{G}_l^m(\vv{k})
+e^{im\va_0}\rrf{\uvk}\frac{z}{z-\mu_0}P_j^m(\mu_0)e^{-im\va_0}.
\label{Eq4b}
\ee
For $|m|\le j\le L$, the above equation can be rewritten as
\be
\sum_{l=|m|}^L\left[\delta_{jl}-c\omega_l^mL_{jl}^m(z)\right]
\bar{G}_l^m(\vv{k})
=\frac{z}{z-\uvk\cdot\uv_0}P_j^m(\uvk\cdot\uv_0)
e^{im\va_0}\rrf{\uvk}e^{-im\va_0}.
\label{Eq5a}
\ee

As we will see below, an exact solution is readily obtained from (\ref{Eq4a}) 
and (\ref{Eq5a}). We introduce the following matrices and vectors.
\ba
\{\vv{L}^m(z)\}_{jl}&=&L_{jl}^m(z),
\\
\{\vv{W}^m\}_{jl}&=&\omega_l^m\delta_{jl},
\\
\{\bar{\vv{G}}^m(\vv{k})\}_l&=&\bar{G}_l^m(\vv{k}),
\\
\{\vv{P}^m(\uvk,\uv)\}_j&=&P_j^m(\uvk\cdot\uv)
e^{im\va_0}\rrf{\uvk}e^{-im\va}.
\ea
We then have
\[
\left[\vv{I}-c\vv{L}^m(z)\vv{W}^m\right]\bar{\vv{G}}^m(\vv{k})
=\frac{z}{z-\uvk\cdot\uv_0}\vv{P}^m(\uvk,\uv_0),
\]
where $\vv{I}$ is the identity. Using (\ref{Eq4a}), $\bar{G}(\vv{k},\uv)$ 
is obtained as
\[
\bar{G}(\vv{k},\uv)
=\frac{z}{z-\uvk\cdot\uv_0}\delta(\uv-\uv_0)
+\frac{c}{4\pi}\frac{z}{z-\uvk\cdot\uv}\sum_{m=-L}^L
{\vv{P}^m(\uvk,\uv)}^{\dagger}\vv{W}^m\bar{\vv{G}}^m(\vv{k})
\]
Therefore we obtain
\ba
G(\vv{r},\uv)
&=&
\frac{e^{-r}}{r^2}\delta(\uv-\hvv{r})\delta(\uv-\uv_0)
\\
&+&
\frac{c}{2(2\pi)^4}\int_{\Rm^3}e^{i\vv{k}\cdot\vv{r}}\frac{z}{z-\uvk\cdot\uv}
\frac{z}{z-\uvk\cdot\uv_0}
\\
&\times&
\sum_{m=-L}^L{\vv{P}^m(\uvk,\uv)}^{\dagger}\vv{W}^m
\left[\vv{I}-c\vv{L}^m(z)\vv{W}^m\right]^{-1}\vv{P}^m(\uvk,\uv_0)\,d\vv{k}.
\ea
Here we used
\be
\int_{\Rm^3}\frac{e^{-r}}{r^2}\delta\left(\uv-\frac{\vv{r}}{r}\right)
e^{-i\vv{k}\cdot\vv{r}}\,d\vv{r}
=\frac{1}{1+i\vv{k}\cdot\uv}.
\label{formula1}
\ee
We note that the first term in the above equation is the uncollided term 
in the collision expansion. By defining
\be
M(\vv{k},\uv,\uv_0)=
\sum_{m=-L}^L{\vv{P}^m(\uvk,\uv)}^{\dagger}\vv{W}^m
\left[\vv{I}-c\vv{L}^m(z)\vv{W}^m\right]^{-1}\vv{P}^m(\uvk,\uv_0),
\label{defM}
\ee
we obtain (\ref{mainmatrix}). If $L=0$, we have
\[
M(\vv{k},\uv,\uv_0)=\frac{1}{1-cL_{00}^0(z)}
=\frac{1}{1-c\frac{i}{k}Q_0(\frac{i}{k})}
=\frac{1}{1-\frac{c}{k}\tan^{-1}(k)},
\]
where we used (\ref{QPz}) and (\ref{Q00}).

\section{Nonstandard Fourier transform}
\label{nonstandard}

Let us explore an alternative formulation suitable for numerical calculation. 
Similarly to the previous section, we will closely follow \cite{Ganapol15}. 
We have
\ba
\left(\uv\cdot\nabla+1\right)\psi(\vv{r},\uv)
&=&
c\sum_{l=0}^L\sum_{m=-l}^l\frac{\beta_l}{2l+1}
\left[\rrf{\uvk}Y_{lm}(\uv)\right]
\int_{\Sm^2}\left[\rrf{\uvk}Y_{lm}^*(\uv')\right]\psi(\vv{r},\uv')\,d\uv'
\\
&+&
\frac{c}{r^2}e^{-r}p(\uv,\uv_0)\delta(\hvv{r}-\uv_0),
\ea
We define
\[
\psi_l^m(\vv{r})=
\sqrt{\frac{4\pi}{2l+1}\frac{(l+m)!}{(l-m)!}}e^{im\va_0}
\int_{\Sm^2}\left[\rrf{\uvk}Y_{lm}^*(\uv)\right]\psi(\vv{r},\uv)\,d\uv.
\]
In the Fourier space we obtain
\bea
\left(1+i\vv{k}\cdot\uv\right)\bar{\psi}(\vv{k},\uv)
&=&
c\sum_{l=0}^L\sum_{m=-l}^l\frac{\omega_l^m}
{\sqrt{4\pi(2l+1)}}\sqrt{\frac{(l+m)!}{(l-m)!}}e^{-im\va_0}
\left(\rrf{\uvk}Y_{lm}(\uv)\right)\bar{\psi}_l^m(\vv{k})
\nonumber \\
&+&
c\frac{p(\uv,\uv_0)}{1+i\vv{k}\cdot\uv_0}.
\label{eq1}
\eea
This is expressed as
\bea
\bar{\psi}(\vv{k},\uv)
&=&
\frac{c}{2}\frac{1}{1+i\vv{k}\cdot\uv}\sum_{l=0}^L\sum_{m=-l}^l
\frac{\omega_l^m}{\sqrt{\pi(2l+1)}}\sqrt{\frac{(l+m)!}{(l-m)!}}e^{-im\va_0}
\left(\rrf{\uvk}Y_{lm}(\uv)\right)\bar{\psi}_l^m(\vv{k})
\nonumber \\
&+&
c\frac{p(\uv,\uv_0)}{(1+i\vv{k}\cdot\uv)(1+i\vv{k}\cdot\uv_0)}.
\label{Eq4apsi}
\eea
We obtain
\ba
\bar{\psi}_j^m(\vv{k})
&=&
\Theta(L-|m|)c\sum_{l=|m|}^L\omega_l^mL_{jl}^m(z)
\\
&\times&
\left[\bar{\psi}_l^m(\vv{k})+\frac{1}{\sqrt{(2j+1)(2l+1)}}
e^{im\va_0}\rrf{\uvk}\frac{z}{z-\mu_0}P_l^m(\mu_0)e^{-im\va_0}\right].
\ea
For $|m|\le j\le L$, the above equation can be rewritten as
\be
\sum_{l=|m|}^L\left[\delta_{jl}-c\omega_l^mL_{jl}^m(z)\right]
\bar{\psi}_l^m(\vv{k})
=\sum_{l=|m|}^L\frac{c\omega_l^mL_{jl}^m(z)}{\sqrt{(2j+1)(2l+1)}}
\frac{z}{z-\uvk\cdot\uv_0}P_l^m(\uvk\cdot\uv_0)
e^{im\va_0}\rrf{\uvk}e^{-im\va_0}.
\label{Eq5apsi}
\ee
As is calculated in Appendix \ref{fouriersubtr}, an expression of 
$\psi(\vv{r},\uv)$ similar to (\ref{mainmatrix}) is obtained using 
(\ref{Eq4apsi}) and (\ref{Eq5apsi}).

Let us look at (\ref{eq1}). By using 
$(2l+1)\mu P_l^m(\mu)=(l-m+1)P_{l+1}^m(\mu)+(l+m)P_{l-1}^m(\mu)$, we have
\[
\mu Y_{lm}(\uv)=\sqrt{[(l+1)^2-m^2]/[4(l+1)^2-1]}Y_{l+1,m}(\uv)
+\sqrt{[l^2-m^2]/[4l^2-1]}Y_{l-1,m}(\uv).
\]
Therefore,
\be
zh_l\bar{\psi}_l^m(\uvk)-(l+1-m)\bar{\psi}_{l+1}^m(\uvk)
-(l+m)\bar{\psi}_{l-1}^m(\uvk)
=zS_l^m,\qquad l>|m|,
\label{Eq10a}
\ee
and
\[
zh_{|m|}\bar{\psi}_{|m|}^m(\uvk)-(|m|+1-m)\bar{\psi}_{|m|+1}^m(\uvk)
=zS_{|m|}^m,
\]
where
\[
h_l=2l+1-\Theta(L-l)c\omega_l^m\frac{(l+m)!}{(l-m)!}
=2l+1-\Theta(L-l)c\beta_l,
\]
and
\ba
S_l^m
&=&
\Theta(L-l)\frac{c\beta_lz}{z-\uvk\cdot\uv_0}
e^{im\va_0}\left[\rrf{\uvk}P_l^m(\mu_0)e^{-im\va_0}\right]
\\
&=&
\Theta(L-l)\frac{c\beta_lz}{z-\uvk\cdot\uv_0}
\sum_{m'=-l}^le^{-im'\va_{\uvk}}d_{m'm}^l(\theta_{\uvk})P_l^{m'}(\mu_0)
e^{i(m-m')\va_0}.
\ea

We introduce Chandrasekhar polynomials of the first and second kinds as
\ba
&&
(l+m)g_{l-1}^m(z)-zh_lg_l^m(z)+(l+1-m)g_{l+1}^m(z)=0,
\\
&&
g_m^m(z)=\frac{(2m)!}{2^mm!}=(2m-1)!!,\quad g_{m+1}^m(z)=zh_mg_m^m(z),
\ea
and
\ba
&&
(l+m)\rho_{l-1}^m(z)-zh_l\rho_l^m(z)+(l+1-m)\rho_{l+1}^m(z)=0,
\\
&&
\rho_m^m(z)=0,\quad \rho_{m+1}^m(z)=\frac{z}{(2m-1)!!},
\ea
where $0\le m\le l$. We note that
\[
g_l^{-m}(z)=(-1)^m\frac{(l-m)!}{(l+m)!}g_l^m(z),\qquad
\rho_l^{-m}(z)=(-1)^m\frac{(l-m)!}{(l+m)!}\rho_l^m(z).
\]
We can express $\bar{\psi}_l^m$ as
\be
\bar{\psi}_l^m=a^m(z)g_l^m(z)+b^m(z)\rho_l^m(z)+\left(1-\delta_{l,|m|}\right)
z\sum_{j=|m|+1}^l\alpha_{l,j}^m(z)S_j^m.
\label{Eq11}
\ee
By setting $l=|m|$ in (\ref{Eq11}), we first notice that
\[
a^m(z)=\frac{2^mm!}{(2m)!}\bar{\psi}_m^m,\qquad
a^{-m}(z)=(-1)^m2^mm!\bar{\psi}_m^{-m},
\]
for $m\ge0$. By plugging (\ref{Eq11}) we have
\[
-(|m|+1-m)\left[b^m(z)\rho_{|m|+1}^m(z)+z\alpha_{|m|+1,|m|+1}^m(z)S_{|m|+1}^m
\right]=zS_{|m|}^m.
\]
Suppose $l>m$. Let us impose
\be
zh_l\alpha_{l,j}^m-(l+1-m)\alpha_{l+1,j}^m-(l+m)\alpha_{l-1,j}^m=0.
\label{Eq14c}
\ee
By substituting (\ref{Eq11}) for $\bar{\psi}_l^m$ in (\ref{Eq10a}), we 
obtain
\[
zh_l\alpha_{l,l}^mS_l^m-(l+1-m)\left(\alpha_{l+1,l}^mS_l^m
+\alpha_{l+1,l+1}^mS_{l+1}^m\right)=S_l^m.
\]
The left-hand side of the above equation can be rewritten as
\[
\mbox{LHS}=-(l+1-m)\alpha_{l+1,l+1}^mS_{l+1}^m+(l+m)\alpha_{l-1,l}^mS_l^m.
\]
Hence we can put
\be
\alpha_{l-1,l}^m=\frac{1}{l+m},\qquad\alpha_{l+1,l+1}^m=0.
\label{Eq15}
\ee
Thus we find
\ba
b^m(z)&=&-(2m-1)!!S_m^m,
\\
b^{-m}(z)&=&-(-1)^m(2m)!(2m-1)!!S_m^{-m},
\ea
for $m\ge0$. To find $\alpha_{l,j}^m(z)$, let us plug the expression 
$\alpha_{l,j}^m=u_j^mg_l^m+v_j^m\rho_l^m$ into $\alpha_{l,l}^m=0$ and 
$\alpha_{l-1,l}^m=1/(l+m)$. We obtain
\ba
u_l^m&=&\frac{\rho_l^m}{(l+m)[g_{l-1}^m\rho_l^m-g_l^m\rho_{l-1}^m]},
\\
v_l^m&=&\frac{-g_l^m}{(l+m)[g_{l-1}^m\rho_l^m-g_l^m\rho_{l-1}^m]}.
\ea
Since we have the relation (Appendix \ref{CD})
\[
(l+m)[g_{l-1}^m(z)\rho_l^m(z)-g_l^m(z)\rho_{l-1}^m(z)]
=\frac{(l+m)!}{(l-m)!}\frac{z}{(2|m|)!},
\]
we obtain
\[
z\alpha_{l,j}^m(z)=\frac{(l-m)!(2|m|)!}{(l+m)!}
\left[\rho_j^m(z)g_l^m(z)-g_j^m(z)\rho_l^m(z)\right].
\]
Finally for $-l\le m\le l$, (\ref{Eq11}) becomes
\bea
\bar{\psi}_l^m
&=&
\frac{g_l^m(z)}{g_{|m|}^m(z)}\bar{\psi}_{|m|}^m
-z\frac{\rho_l^m(z)}{\rho_{|m|+1}^m(z)}S_{|m|}^m
\nonumber \\
&+&
\left(1-\delta_{l,|m|}\right)\frac{(l-m)!(2|m|)!}{(l+m)!}\sum_{j=|m|+1}^l
\left[\rho_j^m(z)g_l^m(z)-g_j^m(z)\rho_l^m(z)\right]S_j^m
\nonumber \\
&=&
\frac{g_l^m(z)}{g_{|m|}^m(z)}\bar{\psi}_{|m|}^m
+\frac{\chi_l^m(z)}{g_{|m|}^m(z)},
\label{Eq22a}
\eea
where
\bea
&&
\chi_l^m(z)
=g_{|m|}^m(z)
\nonumber \\
&&
\times\left\{
\left(1-\delta_{l,|m|}\right)\frac{(l-m)!(2|m|)!}{(l+m)!}\sum_{j=|m|+1}^l
\left[\rho_j^m(z)g_l^m(z)-g_j^m(z)\rho_l^m(z)\right]S_j^m
-z\frac{\rho_l^m(z)}{\rho_{|m|+1}^m(z)}S_{|m|}^m\right\}.
\nonumber \\
\label{defchi}
\eea

To find the initial term $\bar{\psi}_{|m|}^m$, we set $j=|m|$ in 
(\ref{Eq5a}) and obtain
\ba
\sum_{l=|m|}^L\left[\delta_{|m|,l}-c\omega_l^mL_{|m|,l}^m(z)\right]
\bar{\psi}_l^m
&=&
\frac{cz}{z-\uvk\cdot\uv_0}\sum_{l=|m|}^L
\frac{\omega_l^mL_{|m|,l}^m(z)}{\sqrt{(2|m|+1)(2l+1)}}
\\
&\times&
P_l^m(\uvk\cdot\uv_0)e^{im\va_0}\rrf{\uvk}e^{-im\va_0}.
\ea
We can write
\be
L_{|m|,l}^m
=\frac{z}{2}\int_{-1}^1\frac{P_{|m|}^m(\mu)P_l^m(\mu)}{z-\mu}\,d\mu
=zQ_l^m(z)P_{|m|}^m(z),
\label{QPz}
\ee
where $P_l^m(z)$ and $Q_l^m(z)$ are analytically continued associated 
Legendre functions of the first and second kinds which have a branch cut 
from $-\infty$ to $1$. They satisfy the recurrence relation 
(\ref{recurrencePlm}) with initial terms
\[
P_m^m(z)=(2m-1)!!(z-1)^{m/2}(z+1)^{m/2},\quad
P_{m+1}^m(z)=(2m+1)z P_m^m(z),
\]
and
\[
Q_m^m(z)=\frac{[(2m-1)!!]^2}{2P_m^m(z)}
\int_{-1}^1\frac{(1-\mu^2)^m}{z-\mu}\,d\mu,\quad
Q_{m+1}^m(z)=(2m+1)zQ_m^m(z)-\frac{(2m)!}{P_m^m(z)}.
\]
For $|m|\le L$, we obtain
\bea
\bar{\psi}_{|m|}^m
&=&
\frac{cz}{\Lambda^m(z)}\sum_{l=|m|}^L\omega_l^mQ_l^m(z)P_{|m|}^m(z)
\nonumber \\
&\times&
\left[\frac{1}{\sqrt{(2|m|+1)(2l+1)}}\frac{z}{z-\uvk\cdot\uv_0}
P_l^m(\uvk\cdot\uv_0)e^{im\va_0}\rrf{\uvk}e^{-im\va_0}
+\frac{\chi_l^m(z)}{g_{|m|}^m(z)}\right],
\nonumber \\
\label{Eq23a}
\eea
where (Appendix \ref{CD})
\ba
\Lambda^m(z)
&=&
1-cz\sum_{l=|m|}^L\omega_l^mQ_l^m(z)P_{|m|}^m(z)\frac{g_l^m(z)}{g_{|m|}^m(z)}
\\
&=&
\frac{(L+1-m)!}{(L+m)!}\frac{P_{|m|}^m(z)}{g_{|m|}^m(z)}
\left[g_{L+1}^m(z)Q_L^m(z)-g_L^m(z)Q_{L+1}^m(z)\right].
\ea
We can calculate $\bar{\psi}_l^m(\vv{k})$ using 
(\ref{Eq22a}) and (\ref{Eq23a}).

We note that the Fourier transform of the angular flux is given by
\[
\bar{\psi}(\vv{k},\uv)
=\sum_{l=0}^{\infty}\sum_{m=-l}^l
\sqrt{\frac{2l+1}{4\pi}\frac{(l-m)!}{(l+m)!}}e^{-im\va_0}
\bar{\psi}_l^m(\vv{k})\rrf{\uvk}Y_{lm}(\uv).
\]
Using (\ref{Eq22a}) the above equation is rewritten as
\[
\bar{\psi}(\vv{k},\uv)
=\sum_{m=-\infty}^{\infty}\left[
\phi_{\hvv{k}}^m(z,\uv)\bar{\psi}_{|m|}^m(z,\hvv{k})+T_{\hvv{k}}^m(z,\uv)
\right],
\]
where
\ba
\phi_{\hvv{k}}^m(z,\uv)&=&
\frac{1}{g_{|m|}^m(z)}\sum_{l=|m|}^{\infty}\sqrt{\frac{2l+1}{4\pi}
\frac{(l-m)!}{(l+m)!}}e^{-im\va_0}g_l^m(z)\rrf{\uvk}Y_{lm}(\uv),
\\
T_{\hvv{k}}^m(z,\uv)&=&
\frac{1}{g_{|m|}^m(z)}\sum_{l=|m|}^{\infty}\sqrt{\frac{2l+1}{4\pi}
\frac{(l-m)!}{(l+m)!}}e^{-im\va_0}\chi_l^m(z)\rrf{\uvk}Y_{lm}(\uv).
\ea
Note that the dependence of $\bar{\psi}_m^m$ in (\ref{Eq23a}) on $\vv{k}$ is 
split into $z$ and $\hvv{k}$. The angular flux is then given by
\bea
\psi(\vv{r},\uv)
&=&
\frac{1}{(2\pi)^3}\int_{\Rm^3}e^{i\vv{k}\cdot\vv{r}}
\bar{\psi}(\vv{k},\uv)\,d\vv{k}
\nonumber \\
&=&
\frac{1}{(2\pi)^3}\int_{\Sm^2}\int_0^{\infty}k^2e^{ikr\uvk\cdot\hvv{r}}
\sum_{m=-\infty}^{\infty}\left[
\phi_{\hvv{k}}^m(i/k,\uv)\bar{\psi}_{|m|}^m(i/k,\hvv{k})+T_{\hvv{k}}^m(i/k,\uv)
\right]\,dkd\hvv{k}.
\nonumber \\
\label{mainresult0}
\eea
By explicitly writing (\ref{mainresult0}), we obtain (\ref{mainresult}). 

We have
\ba
\int_0^{2\pi}e^{ikr\uvk\cdot\hvv{r}}e^{-im'\va_{\uvk}}\,d\va_{\uvk}
&=&
\int_0^{2\pi}
e^{ikr\sin\theta_{\uvk}\sin\theta_{\hvv{r}}\cos(\va_{\uvk}-\va_{\hvv{r}})}
e^{ikr\cos\theta_{\uvk}\cos\theta_{\hvv{r}}}e^{-im'\va_{\uvk}}\,d\va_{\uvk}
\\
&=&
e^{ikr\cos\theta_{\uvk}\cos\theta_{\hvv{r}}}e^{-im'\va_{\hvv{r}}}
\int_0^{2\pi}e^{ikr\sin\theta_{\uvk}\sin\theta_{\hvv{r}}\cos\va_{\uvk}}
e^{-im'\va_{\uvk}}\,d\va_{\uvk}
\\
&=&
2\pi i^{m'}J_{m'}\left(kr\sin\theta_{\uvk}\sin\theta_{\hvv{r}}\right)
e^{ikr\cos\theta_{\uvk}\cos\theta_{\hvv{r}}}e^{-im'\va_{\hvv{r}}},
\ea
where we noted the Hansen-Bessel formula:
\be
J_m(x)=\frac{1}{2\pi i^m}\int_0^{2\pi}e^{ix\cos\va}e^{-im\va}\,d\va.
\label{formula2}
\ee
Hence we have (\ref{mainresult2}) if $\bar{\psi}_{|m|}^m$ does not depend 
on $\va_{\uvk}$.

\section{Energy density}
\label{numerics}

We consider the energy density $u(\vv{r})$ using (\ref{mainresult2}). 
Let us assume an isotropic source $\delta(\vv{r})$. Without loss of 
generality we can set
\[
\va_{\hvv{r}}=0,\qquad\theta_{\hvv{r}}=0.
\]
We normalize $u(\vv{r})$ with the speed of neutrons. We compute $u(\vv{r})$ as
\bea
u(\vv{r})
&=&
\int_{\Sm^2}\int_{\Sm^2}G(\vv{r},\uv;\uv_0)\,d\uv d\uv_0
\nonumber \\
&=&
\frac{1}{r^2}e^{-r}+
\frac{1}{(2\pi)^2}\int_0^{\infty}k^2\int_{-1}^1e^{ikr\mu_{\uvk}}\int_{\Sm^2}
\kappa_{00}(\vv{k})\,d\uv_0\,d\mu_{\uvk}dk.
\label{density}
\eea
We note that $\kappa_{00}(\vv{k})$ is obtained as
\[
\kappa_{00}(\vv{k})=
\left\{(L+1)\left[g_{L+1}^0(\frac{i}{k})Q_L^0(\frac{i}{k})
-g_L^0(\frac{i}{k})Q_{L+1}^0(\frac{i}{k})\right]\right\}^{-1}
\frac{c\frac{i}{k}Q_0(\frac{i}{k})}{1+i\vv{k}\cdot\uv_0},
\]
where
\be
Q_0\left(\frac{i}{k}\right)=\frac{1}{2}\ln\frac{i/k+1}{i/k-1}=-i\tan^{-1}(k).
\label{Q00}
\ee
Noting that
\[
\int_{\Sm^2}\frac{1}{1+i\vv{k}\cdot\uv_0}\,d\uv_0
=4\pi\frac{\tan^{-1}(k)}{k},
\]
we obtain
\[
u(\vv{r})
=\frac{1}{r^2}e^{-r}+
\frac{c}{\pi(L+1)}\int_0^{\infty}
\frac{2iQ_0(\frac{i}{k})\sin(kr)\tan^{-1}(k)}
{kr\left[g_{L+1}^0(\frac{i}{k})Q_L(\frac{i}{k})
-g_L^0(\frac{i}{k})Q_{L+1}(\frac{i}{k})\right]}\,dk.
\]
In particular if $L=0$ (isotropic scattering), we have
\[
u(\vv{r})
=\frac{1}{r^2}e^{-r}+
\frac{2c}{\pi}\int_0^{\infty}
\frac{\sin(kr)[\tan^{-1}(k)]^2}
{kr\left[1-\frac{c}{k}\tan^{-1}(k)\right]}\,dk.
\]
This is the expression obtained from the textbook Fourier transform 
approach shown in Appendix \ref{textbook}.

\section{Concluding remarks}

We have obtained the angular flux for anisotropic scattering by means of the 
Fourier transform. The calculation developed here can be compared to the 
method of rotated reference frames \cite{Markel04,Panasyuk06}, which uses the 
Fourier transform and rotated reference frames. 
It is known that the method of rotated reference frames has instability 
when spherical haromonics with large degrees are used \cite{Markel}. 
In the present formulation, decomposition of the angular flux into eigenmodes 
is not introduced. In this way, such instability is removed.

\section*{Acknowledgement}

The main results of this paper were obtained by extending the 
one-dimensional Fourier transform calculation by B. D. Ganapol. The author 
learned this calculation at the 24th International Conference on Transport 
Theory (ICTT24) held in Taormina, Sicily, Italy in September 2015.

\appendix

\section{The case of isotropic scattering}
\label{textbook}

In the isotropic case we can obtain $G(\vv{r},\uv;\uv_0)$ as follows 
\cite{Case-Zweifel,Ishimaru78}. The Green's function obeys
\[
\uv\cdot\nabla G(\vv{r},\uv;\uv_0)+G(\vv{r},\uv;\uv_0)=
\frac{c}{4\pi}\int_{\Sm^2}G(\vv{r},\uv;\uv_0)\,d\uv
+\delta(\vv{r})\delta(\uv-\uv_0).
\label{rte}
\]
We write $G$ in terms of $G_0$ as
\ba
G(\vv{r},\uv;\uv_0)
&=&
\int_{\Rm^3}\int_{\Sm^2}G_0(\vv{r}-\vv{r}',\uv;\uv')
\\
&\times&
\left\{\frac{c}{4\pi}\int_{\Sm^2}G(\vv{r}',\uv'';\uv_0)
\,d\uv''+\delta(\vv{r}')\delta(\uv'-\uv_0)\right\}\,d\vv{r}'d\uv'.
\ea
We define
\ba
U(\vv{r};\uv_0)
&=&
\int_{\Sm^2}G(\vv{r},\uv'';\uv_0)\,d\uv''
\\
&=&
\int_{\Rm^3}\frac{1}{|\vv{r}-\vv{r}'|^2}e^{-|\vv{r}-\vv{r}'|}
\left\{\frac{c}{4\pi}U(\vv{r}';\uv_0)+\delta(\vv{r}')
\delta\left(\frac{\vv{r}-\vv{r}'}{|\vv{r}-\vv{r}'|}-\uv_0\right)
\right\}\,d\vv{r}'.
\ea
Since the Fourier transform is obtained as
\[
\bar{U}(\vv{k})=
\frac{1}{1+i\vv{k}\cdot\uv_0}\left[1-\frac{c}{k}\tan^{-1}(k)\right]^{-1},
\]
we obtain
\[
U(\vv{r};\uv_0)=
\frac{1}{(2\pi)^3}\int_{\Rm^3}
\frac{e^{i\vv{k}\cdot\vv{r}}}{1+i\vv{k}\cdot\uv_0}
\left[1-\frac{c}{k}\tan^{-1}(k)\right]^{-1}
\,d\vv{k}.
\]
Finally, the Green's function is written as
\ba
G(\vv{r},\uv;\uv_0)
&=&
G_0(\vv{r},\uv;\uv_0)
+\frac{c}{4\pi(2\pi)^3}\int_{\Rm^3}e^{i\vv{k}\cdot\vv{r}}
\nonumber \\
&\times&
\frac{1}{(1+i\vv{k}\cdot\uv)(1+i\vv{k}\cdot\uv_0)}
\left[1-\frac{c}{k}\tan^{-1}(k)\right]^{-1}\,d\vv{k}.
\ea

Let us consider the energy density for the isotropic source $\delta(\vv{r})$. 
We obtain
\ba
u(\vv{r})
&=&
\int_{\Sm^2\times\Sm^2}G(\vv{r},\uv;\uv')\,d\uv d\uv'
\\
&=&
\int_{\Sm^2\times\Sm^2}G_0(\vv{r},\uv;\uv')\,d\uv d\uv'
+
\frac{c}{4\pi(2\pi)^3}\int_{\Rm^3}e^{i\vv{k}\cdot\vv{r}}
\\
&\times&
\int_{\Sm\times\Sm^2}\frac{1}{(1+i\vv{k}\cdot\uv)(1+i\vv{k}\cdot\uv')}
\,d\uv d\uv'
\left[1-\frac{c}{k}\tan^{-1}(k)\right]^{-1}\,d\vv{k}.
\\
&=&
\frac{e^{-r}}{r^2}+\frac{2c}{\pi}\int_0^{\infty}\frac{\sin(kr)}{r}
\frac{\left[\tan^{-1}(k)\right]^2}{k-c\tan^{-1}(k)}\,dk.
\ea

\section{Christoffel-Darboux formulas}
\label{CD}

We consider the following two recurrence relations \cite{Ganapol15}.
\bea
za_lq_l^m(z)-(l+1-m)q_{l+1}^m(z)-(l+m)q_{l-1}^m(z)&=&0,
\label{Eq27a}
\\
\mu b_lr_l^m(\mu)-(l+1-m)r_{l+1}^m(\mu)-(l+m)r_{l-1}^m(\mu)&=&0.
\label{Eq27b}
\eea
By subtracting $q_l^m(z)[(l-m)!/(l+m)!]\times(\ref{Eq27b})$ from 
$r_l^m(\mu)[(l-m)!/(l+m)!]\times(\ref{Eq27a})$ 
we obtain \cite{Garcia-Siewert82}
\be
(za_l-\mu b_l)\frac{(l-m)!}{(l+m)!}q_l^m(z)r_l^m(z)+t_{l+1}^m(z,\mu)
-t_l^m(z,\mu)=0,
\ee
where
\[
t_l^m(z,\mu)=\frac{(l-m)!}{(l-1+m)!}\left[q_{l-1}^m(z)r_l^m(\mu)
-q_l^m(z)r_{l-1}^m(\mu)\right].
\]
Suppose $l_0>|m|+1$.  By taking the summation $\sum_{l=|m|+1}^{l_0}$ we obtain
\ba
&&
\frac{(l_0+1-m)!}{(l_0+m)!}
\left[q_{l_0}^m(z)r_{l_0+1}^m(\mu)-q_{l_0+1}^m(z)r_{l_0}^m(\mu)\right]
\\
&&
=\sum_{l=|m|+1}^{l_0}\left(\mu b_l-za_l\right)\frac{(l-m)!}{(l+m)!}
q_l^m(z)r_l^m(\mu)
\\
&&
+\frac{(|m|+1-m)!}{(|m|+m)!}
\left[q_{|m|}^m(z)r_{|m|+1}^m(\mu)-q_{|m|+1}^m(z)r_{|m|}^m(\mu)\right].
\ea

If we set
\[
l_0=l-1,\quad a_l=b_l=h_l,\quad q_l^m=g_l^m,\quad r_l^m=\rho_l^m,
\]
we obtain
\[
(l+m)\left[g_{l-1}^m(z)\rho_l^m(z)-g_l^m(z)\rho_{l-1}^m(z)\right]
=\frac{(l+m)!}{(l-m)!}\frac{z}{(2|m|)!}.
\]

If we set
\[
l_0=L,\quad a_l=h_l,\quad b_l=2l+1,\quad q_l^m=\frac{g_l^m}{g_{|m|}^m},\quad 
r_l^m=Q_l^mP_{|m|}^m,
\]
we obtain
\ba
&&
\frac{(L+1-m)!}{(L+m)!}\frac{P_{|m|}^m(z)}{g_{|m|}^m(z)}
\left[g_{L+1}^m(z)Q_L^m(z)-g_L^m(z)Q_{L+1}^m(z)\right]
\\
&&
=-cz\sum_{l=|m|+1}^L\omega_l^mg_l^m(z)Q_l^m(z)
\\
&&
-\frac{(|m|+1-m)!}{(|m|+m)!}\frac{P_{|m|}^m(z)}{g_{|m|}^m(z)}
\left[g_{|m|}^m(z)Q_{|m|+1}^m(z)-g_{|m|+1}^m(z)Q_{|m|}^m(z)\right].
\ea
Therefore we have
\ba
&&
\frac{(L+1-m)!}{(L+m)!}\frac{P_{|m|}^m(z)}{g_{|m|}^m(z)}
\left[g_{L+1}^m(z)Q_L^m(z)-g_L^m(z)Q_{L+1}^m(z)\right]
\\
&&
=-cz\sum_{l=|m|}^L\omega_l^m\frac{g_l^m(z)}{g_{|m|}^m(z)}Q_l^m(z)P_{|m|}^m(z)
+1.
\ea
Here we used
\[
Q_{|m|+1}^m(z)=(2|m|+1)zQ_{|m|}^m(z)
-\frac{[(2|m|)!]^{\mathrm{sgn}(m)}}{P_{|m|}^m(z)}.
\]
The above relation is derived from (\ref{QPz}).

\section{The Fourier transform with ballistic subtraction}
\label{fouriersubtr}

We additionally introduce the following matrices and vectors.
\ba
\{\vv{S}\}_{jl}&=&\frac{1}{\sqrt{2j+1}}\delta_{jl},
\\
\{\bar{\vv{\psi}}^m(\vv{k})\}_l&=&\bar{\psi}_l^m(\vv{k}).
\ea
Equation (\ref{Eq5apsi}) can be expressed as
\[
\left[\vv{I}-c\vv{L}^m(z)\vv{W}^m\right]\bar{\vv{\psi}}^m(\vv{k})
=c\vv{S}\vv{L}^m(z)\vv{S}\vv{W}^m(z)\frac{z}{z-\uvk\cdot\uv_0}
\vv{P}^m(\uvk,\uv_0).
\]
Using (\ref{Eq4apsi}), $\bar{\vv{\psi}}^m(\vv{k})$ is obtained as
\[
\bar{\psi}(\vv{k},\uv)
=\frac{cp(\uv,\uv_0)}{(1+i\vv{k}\cdot\uv)(1+i\vv{k}\cdot\uv_0)}
+\frac{c}{4\pi}\frac{z}{z-\uvk\cdot\uv}{\vv{P}^m(\uvk,\uv)}^{\dagger}\vv{W}^m
\bar{\vv{\psi}}^m(\vv{k})
\]
Therefore we obtain
\ba
\psi(\vv{r},\uv)
&=&
cp(\uv,\uv_0)\int_0^{\infty}\int_0^{\infty}e^{-r_1}e^{-r_2}
\delta\left(\vv{r}-r_1\uv-r_2\uv_0\right)\,dr_1dr_2
\\
&+&
\frac{c^2}{2(2\pi)^4}\int_{\Rm^3}e^{i\vv{k}\cdot\vv{r}}\frac{z}{z-\uvk\cdot\uv}
\frac{z}{z-\uvk\cdot\uv_0}
\\
&\times&
{\vv{P}^m(\uvk,\uv)}^{\dagger}\vv{W}^m
\left[\vv{I}-c\vv{L}^m\vv{W}^m\right]^{-1}\vv{S}\vv{L}^m(z)\vv{S}\vv{W}^m(z)
\vv{P}^m(\uvk,\uv_0)\,d\vv{k}.
\ea
We note that the first term in the above equation is the once-collided term 
in the collision expansion:
\[
(\mbox{once-collided term})=cp(\uv,\uv_0)
\Theta(\pi-\tau-\tau_0)\delta(|\va-\va_0|-\pi)
\frac{e^{-r(\sin{\tau}+\sin{\tau_0})/\sin(\tau+\tau_0)}}
{r\sin{\tau}\sin{\tau_0}},
\]
where $\cos{\tau}=\hvv{r}\cdot\uv$, $\cos{\tau_0}=\hvv{r}\cdot\uv_0$.


\end{document}